\documentclass[12pt]{article}
\usepackage{graphicx}
\title{Mesons and diquarks in Coulomb-gauge QCD}
\author{R.F. Wagenbrunn%
\\
Institut f\"ur Physik, Fachbereich Theoretische Physik,\\
Karl-Franzens-Universit\"at Graz,
Universit\"atsplatz 5, A-8010 Austria}
\date{}

\begin{document}
\maketitle
\begin{abstract}
I demonstrate how confinement in Coulomb-gauge QCD makes 
quark-quark states of the color anti-triplet (diquarks) move out of the
physical spectrum.
Mesons as color singlet quark-antiquark states, on the other hand
have finite masses and for highly excited states in the meson spectra
effective restoration of chiral symmetry can be observed.
\end{abstract}

In Coulomb-gauge QCD the $00$-component of the gluon propagator
$D_{\mu\nu}(\vec{x},t)$
has an instantaneous part $V_C(\vec{x})\delta(t)$ and confinement means that
$-V_C(\vec{x})\to\infty$ for $|\vec{x}|\to\infty$. It was shown that
Coulomb confinement is a necessary condition for confinement, i.e.,
that the gauge invariant quark-antiquark potential $V_W(r)$ goes to infinity for $r\to\infty$
\cite{Zwanziger:2002sh}.
An almost linearly rising Coulomb potential has been suggested
\cite{conf}, 
which was also confirmed by results from the lattice
\cite{latt}.
In momentum space it becomes
\begin{equation}
V_C(|\vec{k}|)=\frac{\sigma_C}{|\vec{k}|^4},
\label{eq:vcoul}
\end{equation}
where $\sigma_C$ is the Coulomb string tension.
Based on previous works \cite{qcdcoul}
we performed a study of the mechanism of Coulomb confinement in the
Dyson-Schwinger--Bethe-Salpeter framework  \cite{ds} in Rainbow-ladder
approximation \cite{Alkofer:2005ug}.
We took into account only
the Coulomb potential (\ref{eq:vcoul})
and neglected transverse gluons and noninstantaneous contributions
to $D_{00}$. In that way all integrals over
$k_0$ can be performed analytically and 
one has to deal with three-dimensional integral equations only.
(\ref{eq:vcoul}) has also an unrealistic ultraviolet (UV) behavior. However,
it has the advantage that it produces no  UV divergences, thus making renormalization
not necessary. Due to all these approximations some physics is lost and
the model is not expected to provide realistic quantitative results
but some qualitative insight into the physics of confinement in QCD.
Since the axial-vector Ward-Takahashi identity is satisfied,
chiral symmetry and its dynamical breaking
are respected. In particular the pion mass becomes zero in the chiral limit,
i.e., for vanishing current quark mass.
The potential (\ref{eq:vcoul}) causes infrared (IR)
divergences which are regulated by introducing an IR regulator $\mu_{\rm IR}$
and replacing $k^2\to k^2+\mu_{\rm IR}^2$ (here and in the following $k=|\vec{k}|$).
The IR limit is then taken by means of $\mu_{\rm IR}\to 0$.
The essential point is that the integral
\begin{equation}
\frac{1}{2\pi^2}\int d^3q\;\frac{1}{((\vec{p}-\vec{q}|)^2+\mu_{\rm IR}^2)^2}=
\frac{1}{2\mu_{\rm IR}}
\end{equation}
diverges in the IR limit
and one can write
\begin{equation}
\frac{1}{2\pi^2} \int d^3q\;V_C(|\vec{p}-\vec{q}|)f(q)=
{\frac{\sigma_C}{2\mu_{\rm IR}}\int d^3q\;\delta(\vec{p}-\vec{q})f(q)}+ \mbox{IR finite term}.
\label{eq:irlimit}
\end{equation}

\begin{figure}[t]
\includegraphics[height=5.5cm,clip=]{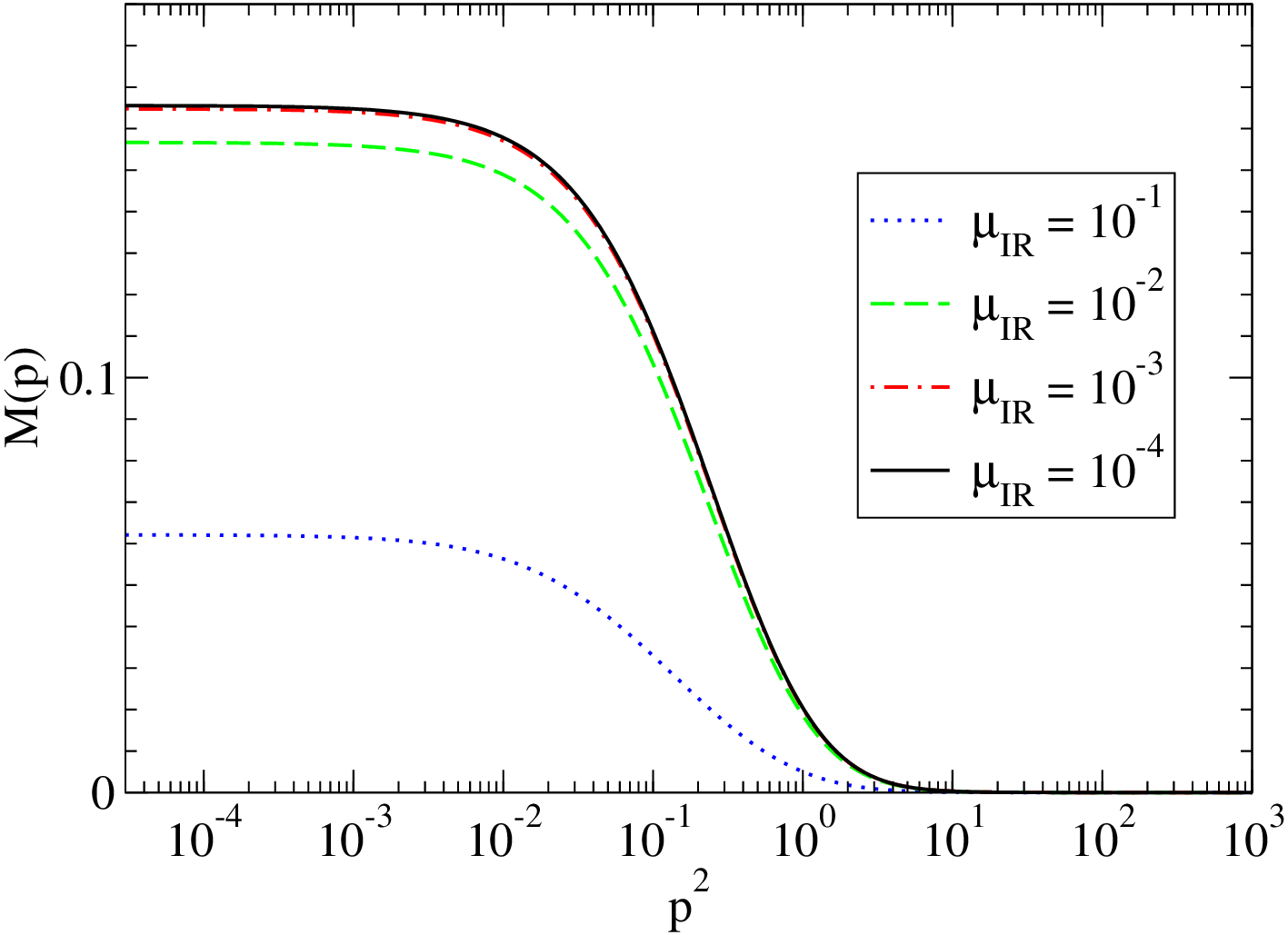}\hfill
\includegraphics[height=5.5cm,clip=]{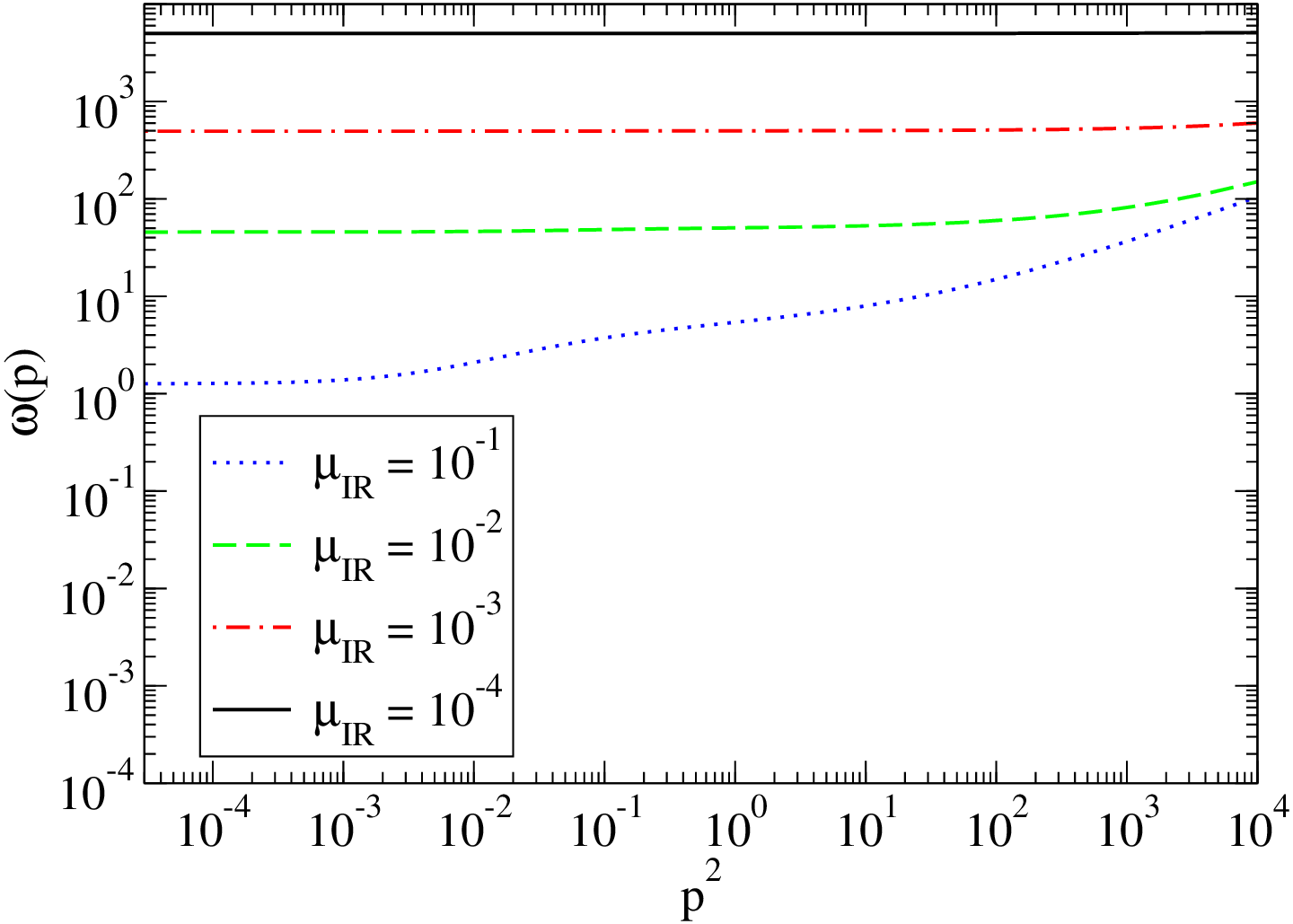}
\includegraphics[height=5.5cm,clip=]{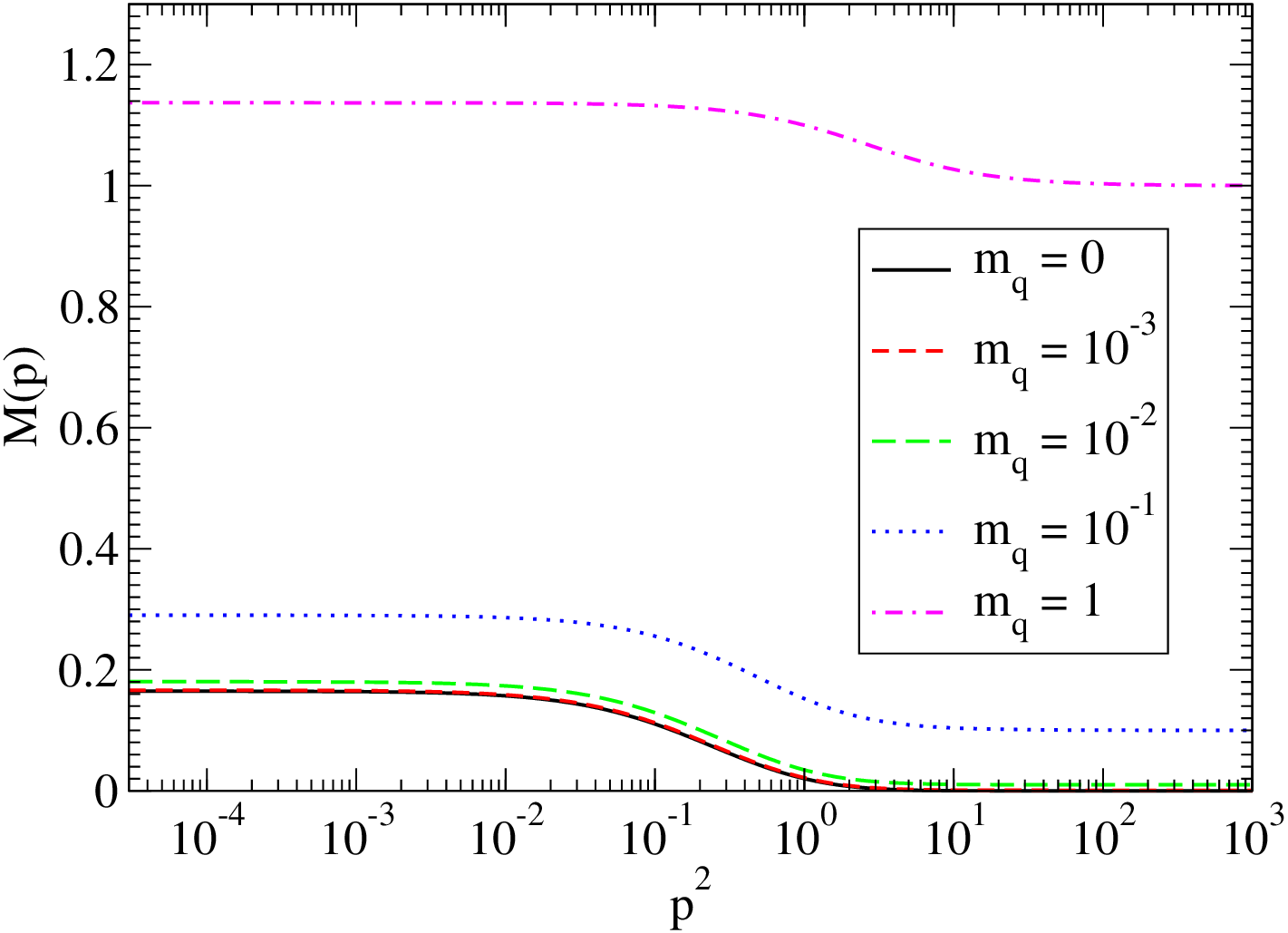}\hfill
\includegraphics[height=5.5cm,clip=]{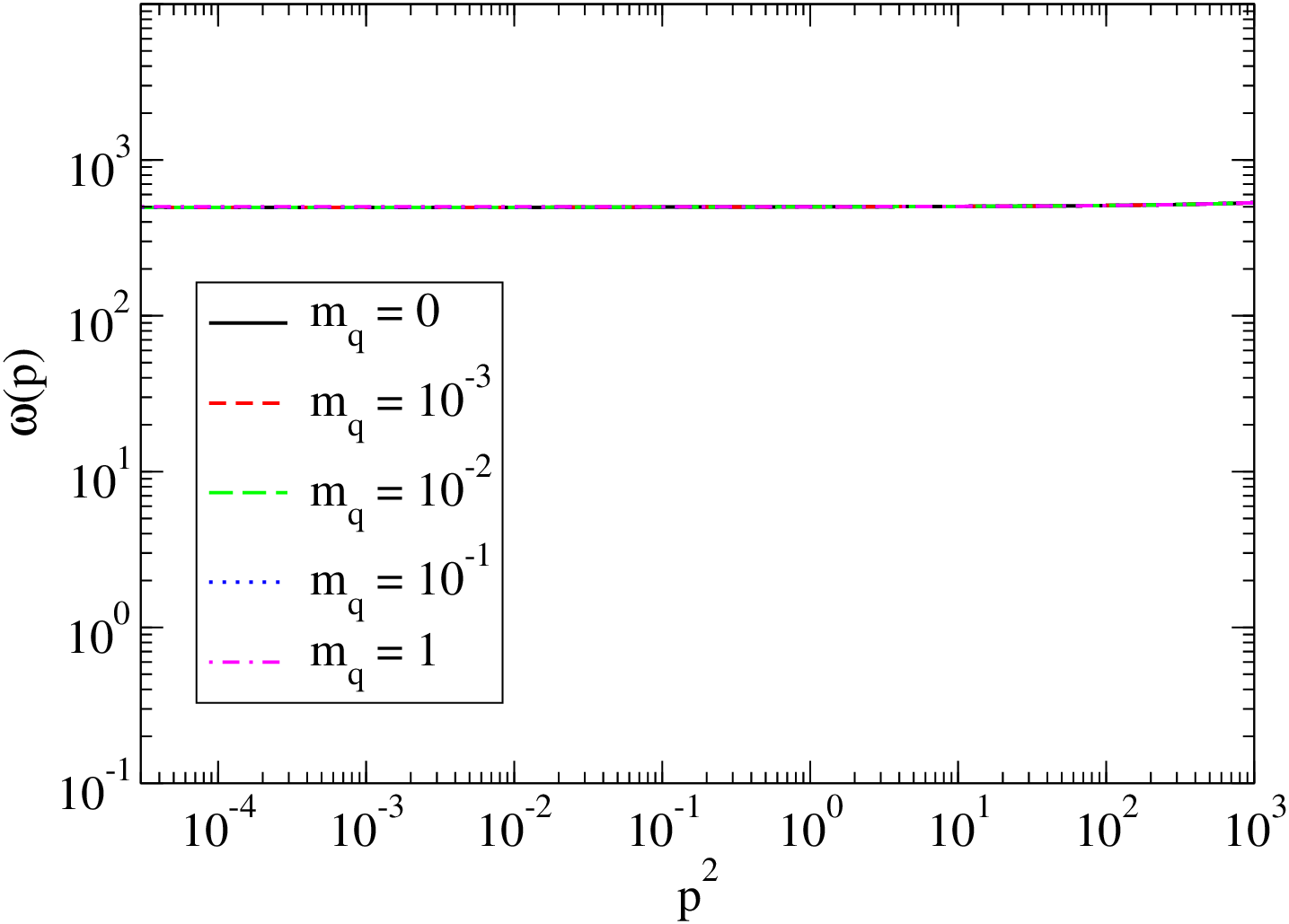}
\caption{Mass function $M(p)$ (left plots) and $\omega(p)$ (right plots): In the
upper plots the IR behavior for current quark mass $m=0$ and in the lower plots
the results for different current quark masses at constant
$\mu_{\rm IR}=10^{-3}\ \sqrt{\sigma_C}$ is shown.
}
\end{figure}
For the gap equation of the quark propagator $S(p)$ the ansatz 
\begin{equation}
S^{-1}(p)=-i\left(\gamma_0p_0-\vec{\gamma}\cdot\vec{p}\;C(p) - B(p)\right)
\end{equation}
leads to a coupled system of two integral equations
\begin{eqnarray}\label{gapequdetaila}
B(p)&=&m+\frac{1}{2\pi^2}\int d^3q\;V_C(|\vec{p}-\vec{q}|)
\frac{M(q)}{\bar{\omega}(q)}\\\label{gapequdetailb}
C(p)&=&1+\frac{1}{2\pi^2}\int d^3q\;V_C(|\vec{p}-\vec{q}|)\;\hat{p}\cdot\hat{q}\;
\frac{q}{p\,\bar{\omega}(q)}\;,
\end{eqnarray}
where $\hat{p}=\vec{p}/p$, $m$ is the 
current-quark mass, $\bar{\omega}(p)=\sqrt{M^2(p)+p^2}$ and 
$M(q)=\frac{B(q)}{C(q)}$ is called the quark mass function.
Using (\ref{eq:irlimit}) yields
\begin{eqnarray}
&&B(p)=\frac{\sigma_C}{2\mu_{\rm IR}}\frac{M(p)}{\bar{\omega}(p)}+ \mbox{IR finite term}
=\frac{\sigma_C}{2\mu_{\rm IR}}\frac{B(p)}{\omega(p)}+ \mbox{IR finite term},\\
&&C(p)=\frac{\sigma_C}{2\mu_{\rm IR}}\frac{1}{\bar{\omega}(p)}+ \mbox{IR finite term}
=\frac{\sigma_C}{2\mu_{\rm IR}}\frac{C(p)}{\omega(p)}+ \mbox{IR finite term},
\end{eqnarray}
with
\begin{equation}
\omega(p)=\sqrt{B^2(p)+p^2C^2(p)})=\frac{\sigma_C}{2\mu_{\rm IR}}+ \mbox{IR finite term}.
\label{eq:omega}
\end{equation}
The functions $B(p)$ and $C(p)$ diverge like $\mu_{\rm IR}^{-1}$ but the mass function
is IR finite.
The IR behavior for $M(p)$ and $\omega(p)$ for $m=0$ is demonstrated in the upper plots of Fig. 1,
while in the lower plots the same quantities for constant
$\mu_{\rm IR}=10^{-3}\ \sqrt{\sigma_C}$ but different current quark masses are shown.
The mass function converges to a finite function. For large $p$ it goes to the current quark
mass and for small $p$ it gets a dynamical
mass which is of approximately the same absolute size for different current quark masses. 
$\omega(p)\mu_{\rm IR}$ indeed becomes a constant $\frac{\sigma_C}{2}$, which is independent
of the current quark mass.

The Bethe-Salpeter equation (BSE) for a meson with mass $M$ is
\begin{equation}
\chi(p,M)= - i\int\frac{d^4q}{(2\pi)^4}V_C(|\vec{p}-\vec{q}|)\;
\gamma_0 S(q_0+M/2,q) \nonumber \\
\chi(q,M)S(q_0-M/2,q)\gamma_0.
\label{GenericSal}
\end{equation}
For the pseudoscalar meson (pion with $M=m_\pi$) the Bethe-Salpeter amplitude $\chi(p,m_\pi)$
contains three (pseudoscalar, axial-vector and tensor) components:
\begin{equation}
\chi(p,m_\pi)={P_p(p)}\gamma_5+m_\pi {P_A(p)}\gamma_0\gamma_5+
m_\pi {P_T(p)}\hat{p}\cdot \vec{\gamma}\gamma_0\gamma_5.
\end{equation}
The BSE is reduced to a coupled system of integral equations
\begin{eqnarray}
\omega(p)h(p)&=&\frac{1}{2\pi^2}\int d^3 q V_C(|\vec{p}-\vec{q}|)
\left(h(q)+\frac{m_\pi^2}{4\omega(q)}g(q)\right),
%
\label{eq:bsetype1a}
\\
\left(\omega(p)-\frac{m_\pi^2}{4\omega(p)}\right)g(p)&=&h(p)
+\frac{1}{2\pi^2}\int d^3 q V_C(|\vec{p}-\vec{q}|)
\frac{M(p)M(q)+\vec{p}\cdot\vec{q}}{\bar{\omega}(p)\bar{\omega}(q)}
g(q)
\label{eq:bsetype1b}
\end{eqnarray}
for the two functions
\begin{equation}
h(p)=\frac{P_p(p)}{\omega(p)},
\end{equation}
\begin{equation}
g(p)=\frac{\omega(p)}{\omega^2(p)-\frac{m_\pi^2}{4}}\left[h(p)+2\frac{M(p)}{\bar{\omega}(p)}P_A(p)
+2\frac{p}{\bar{\omega}(p)}P_T(p)\right].
\end{equation}
The IR behavior of Eqs. (\ref{eq:bsetype1a},\ref{eq:bsetype1b})
follows by using Eq. (\ref{eq:omega}) on the left and Eq. (\ref{eq:irlimit}) on the right
hand sides, respectively, which yields
\begin{eqnarray}
{\frac{\sigma_C}{2\mu_{\rm IR}}h(p)}+\mbox{IR finite term}
&=&{\frac{\sigma_C}{2\mu_{\rm IR}}h(p)}+\mbox{IR finite term},\label{eq:irpiona}
\\
{\frac{\sigma_C}{2\mu_{\rm IR}}g(p)}+\mbox{IR finite term}
&=&{\frac{\sigma_C}{2\mu_{\rm IR}}g(p)}+\mbox{IR finite term}.
\label{eq:irpionb}
\end{eqnarray}
Obviously the IR divergences cancel in both equations and all physical observables, in particular
the mass, are determined by the IR finite terms. The functions $h(p)$ and $g(p)$
have an IR finite limit, too. This is demonstrated for $m=0$ in the left plot
of Fig. 2.

\begin{figure}
\includegraphics[height=5.5cm,clip=]{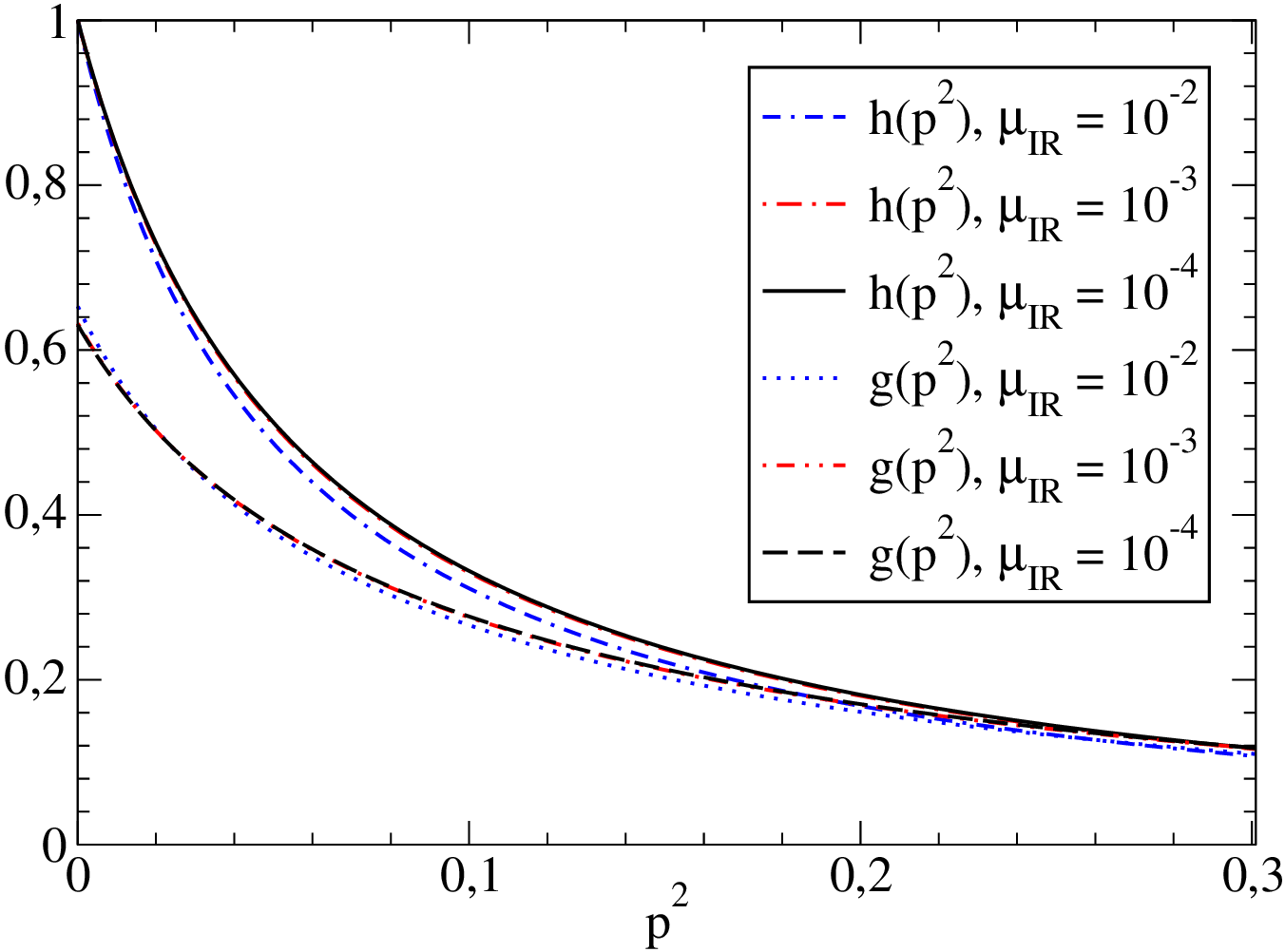}\hfill
\includegraphics[height=5.5cm,clip=]{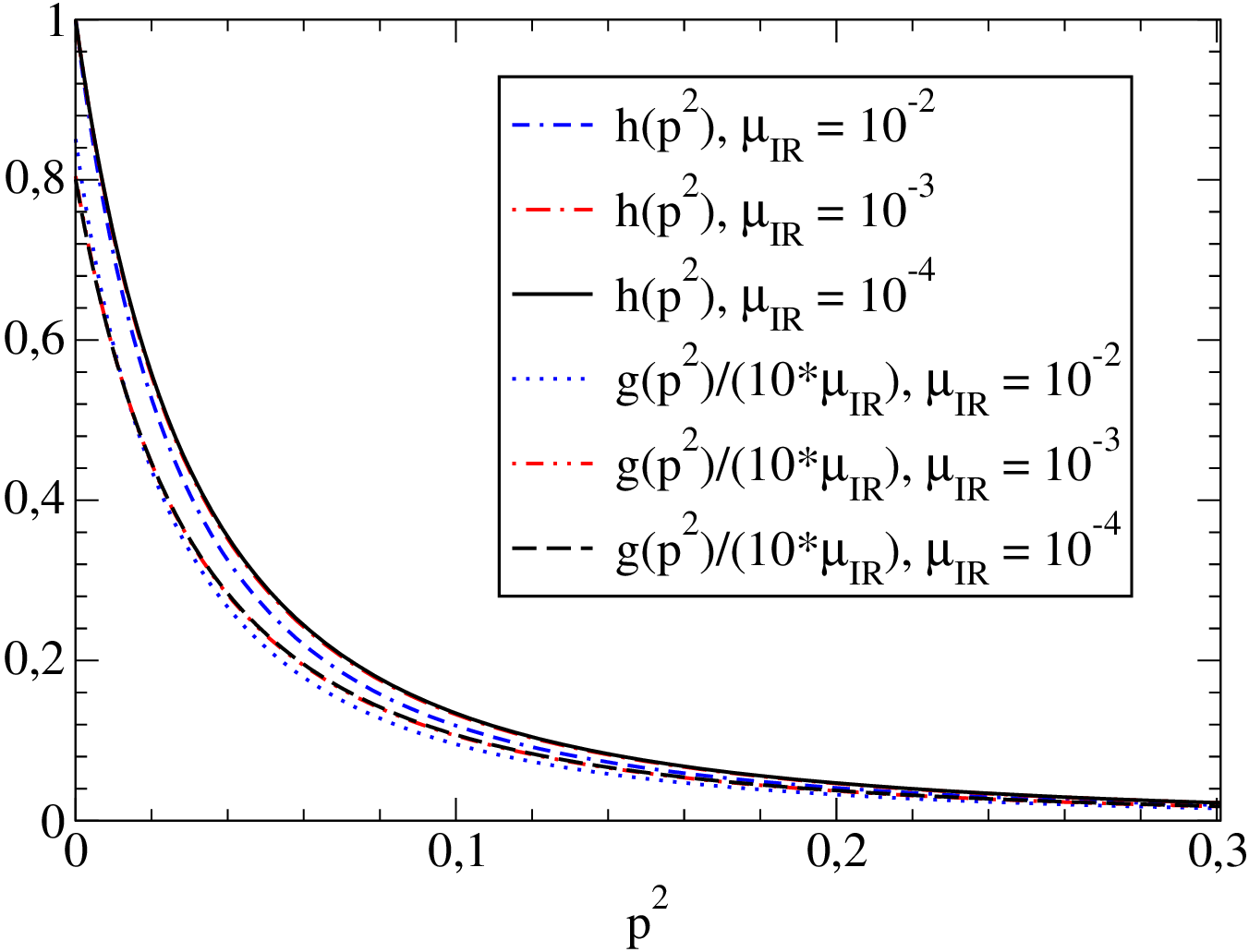}
\caption{IR limit of the functions $h$ and $g$ for the pion (left) and
the scalar diquark (right) for $m=0$. In all cases the normalization has been
chosen such that $h(0)=1$.
}
\end{figure}
For two quarks in the $SU(3)_C$ anti-triplet state (diquark) a color factor $\frac{1}{2}$
enters into the BSE kernel. Apart from that for a scalar diquark with mass $m_{\rm SD}$
one has the same
integral equations as for the pion but
the IR divergent terms on the right hand sides of Eqs. (\ref{eq:irpiona},\ref{eq:irpionb})
are reduced to only half their sizes.
Thus there must be additional
IR divergences. Replacing (\ref{eq:irpiona},\ref{eq:irpionb}) by
\begin{eqnarray}{\frac{\sigma_C}{2\mu_{\rm IR}}}h(p)+\mbox{IR finite term}
={\frac{\sigma_C}{4\mu_{\rm IR}}}h(p)+
\frac{{m_{\rm SD}^2}}{8}g(p)+\mbox{IR finite term},
\\
{\frac{\sigma_C}{2\mu_{\rm IR}}}g(p)-
\frac{{m_{\rm SD}^2}\mu_{\rm IR}}{2\sigma_C}g(p)+\mbox{IR finite term}
=h(p)+{\frac{\sigma_C}{4\mu_{\rm IR}}}g(p)+\mbox{IR finite term}
\end{eqnarray}
and solving for $m_{\rm SD}$ and $g(p)$ yields
\begin{eqnarray}
m_{\rm SD}=\frac{\sigma_C}{2\mu_{\rm IR}},&&
g(p)=\frac{8\mu_{\rm IR}}{\sigma_C}h(p).
\label{eq:sd_ir}
\end{eqnarray}
The divergences are now balanced by introducing a relation
between $h(p)$ and $g(p)$ and making the diquark mass IR divergent. In that way the diquarks are
removed from the physical spectrum. This mechanism of confinement
applies not only for the scalar
diquark but for diquarks of all quantum numbers.
Numerically we have reproduced the IR divergence of the mass
for scalar and axial-vector diquarks \cite{Alkofer:2005ug}. On the other hand, the shape
of the functions $h(p)$ and $g(p)$ converges in the IR limit, which is
demonstrated in the right plot of Fig. 2. For that reason not only
mesons but also diquarks have IR finite radii. Calculations for the electromagnetic
form factor and the charge radius of the the pion in Coulomb-gauge QCD have already
been performed earlier \cite{Langfeld:1989en}.
Explicitly, the charge radius of the pion is given by
\begin{eqnarray}\!\!\!\!\!\!&&\!\!\!\!\!\!\!\!\!\!\!\!
\displaystyle r_\pi^2=\frac{3}{{\cal N}_\pi^2}\int \frac{d^3p}{(2\pi)^3} \left\{
-\frac{3}{32\bar{\omega}^4(p)}\left[2\bar{\omega}^2(p)+\left(M(p)-2p^2M'(p)\right)^2\right]g(p)h(p)
\right.\nonumber \\\!\!\!\!\!\!&&\!\!\!\!\!\!\!\!\!\!\!\!
\displaystyle \left.+\frac{1}{16}\left[g'(p)h(p)+g(p)h'(p)\right]
+\frac{p^2}{24}\left[g(p)h''(p)+g''(p)h(p)-2g'(p)h'(p)\right]
\right\}+{\cal O}(\mu_{\rm IR})
\end{eqnarray}
with
\begin{equation}
{\cal N}_\pi^2=3\int \frac{d^3p}{(2\pi)^3}g(p)h(p),
\end{equation}
and of the scalar diquark by
\begin{eqnarray}
&\displaystyle r_{\rm SD}^2=\frac{3\mu_{\rm IR}}{{\cal N}_{\rm SD}^2}\int \frac{d^3p}{(2\pi)^3} \!\!&\displaystyle \!\!\left\{
-\frac{7}{6\bar{\omega}^4(p)}\left[2\bar{\omega}^2(p)+\left(M(p)-2p^2M'(p)\right)^2\right]h^2(p)
\right.\nonumber \\
&&\displaystyle\; \left.+2h(p)h'(p)+\frac{4p^2}{3}\left[h(p)h''(p)-{h'}^2(p)\right]
\right\}+{\cal O}(\mu_{\rm IR})
\end{eqnarray}
with
\begin{equation}
{\cal N}_{\rm SD}^2=48\mu_{\rm IR}\int \frac{d^3p}{(2\pi)^3}h^2(p).
\end{equation}
In both cases $'$ means $\frac{d}{d(p^2)}$. 
For quark mass $m=0$ we have obtained the results $r_\pi=4.3\ \sigma_C^{-1/2}$ and
$r_{\rm SD}=6.0\ \sigma_C^{-1/2}$ \cite{Alkofer:2005ug}.
Notice that ${\cal N}_\pi^2$ converges to a finite
value while ${\cal N}_{\rm SD}^2$ goes to zero like $\mu_{\rm IR}$. That means that
for the diquark only the shape of $h(p)$ converges but its size diverges like
$\mu_{\rm IR}^{-1/2}$. Due to (\ref{eq:sd_ir}), the size of $g(p)$ goes to zero
like $\mu_{\rm IR}^{1/2}$ on the other hand.

Finally I present results for the highly excited meson spectra in the
chiral limit \cite{Wagenbrunn:2006cs}.
There are certain phenomenological evidences that in highly
excited hadrons the chiral ($SU(2)_L \times SU(2)_R$) and $U(1)_A$
symmetries are approximately restored (for a review see \cite{Glozman:2004gk}).
The states
fall into approximate multiplets of $SU(2)_L \times SU(2)_R$ 
and the mass splittings within the multiplets 
vanish at radial quantum number $n \rightarrow \infty$ and/or spin $ J \rightarrow \infty$.
Furthermore the splittings within a multiplet become much smaller
than between the two subsequent multiplets.
The reason for this ``effective'' symmetry restoration is that
excited hadrons gradually
decouple from the quark condensates due to a diminishing importance of
quantum fluctuations \cite{Glozman:2005tq}. 
I restrict the discussion here to
scalar and pseudoscalar mesons.
Given the complete set of standard quantum numbers $I, J^{PC}$,
the  multiplets of  $SU(2)_L \times SU(2)_R$ for $J=0$ are \cite{Glozman:2003bt}
\begin{eqnarray}
(1/2,1/2)_a   :     1,0^{-+} \longleftrightarrow 0,0^{++}  &\mbox{and}&
(1/2,1/2)_b   :     1,0^{++} \longleftrightarrow 0,0^{-+}  \nonumber.
\end{eqnarray}
The BSE for a scalar meson with mass $m_{0^{++}}$ is reduced to the coupled system of integral equations
\begin{eqnarray}
h(p)\omega(p)=\frac{1}{2\pi^2}\int
d^3q\;V_C(|\vec{p}-\vec{q}|)\;\frac{pq+M(p)M(q)\hat{p}\cdot\hat{q}}{\bar{\omega}(p)\bar{\omega}(q)}
\left(h(q)+\frac{m_{0^{++}}^2}{4\,\omega(q)}g(q)\right),
\label{eq:bsetype2a}\\
g(p)\left(\omega(p)-\frac{m_{0^{++}}^2}{4\,\omega(p)}\right)=
h(p)+\frac{1}{2\pi^2}{\int} d^3q\;V_C(|\vec{p}-\vec{q}|)\hat{p}\cdot\hat{q}g(q).
\label{eq:bsetype2b}
\end{eqnarray}
For highly excited states the typical momenta of the quarks become large.
For large momenta, however, the mass function $M(p)$ becomes small.
Setting $M(p)=0$
in the second equation for the pseudoscalar meson (\ref{eq:bsetype1b})
and the first equation for the scalar meson (\ref{eq:bsetype2a})
gives just the second equation for the scalar meson (\ref{eq:bsetype2b})
and the first equation for the pseudoscalar meson (\ref{eq:bsetype1a}), respectively.
For large momenta with $M(p)\approx 0$ the two systems of coupled integral
equations become approximately the same.
This can explain why pseudoscalar and scalar mesons with large $n$
become approximately degenerate.
For states with $J>0$ similar arguments hold
but there are additional states which fall in the multiplets $(0,0)$ and $(0,1) \oplus (1,0)$.
The  Bethe-Salpeter amplitudes for mesons with
large $J$ become strongly suppressed for small momenta and already
states with $n=0$ become approximately degenerate. In our model the numerical results
for the meson spectra up to $n,J=6$ show 
a very fast restoration of
both $SU(2)_L \times SU(2)_R$ and $U(1)_A$ symmetries with increasing
$J$ and essentially more slow restoration with increasing $n$.
The excited states
lie on approximately linear radial and angular Regge trajectories which is demonstrated
in Fig. 3.
In  the limit $n \rightarrow \infty$ and/or  $J \rightarrow \infty$ 
one observes an approximate degeneracy of all states within the representation
$ [(0,1/2) \oplus (1/2,0)] \times [(0,1/2) \oplus (1/2,0)]$
%
that combines all possible chiral representations
for systems of two massless quarks \cite{Glozman:2003bt}.
\begin{figure}
\includegraphics[height=5.5cm,clip=]{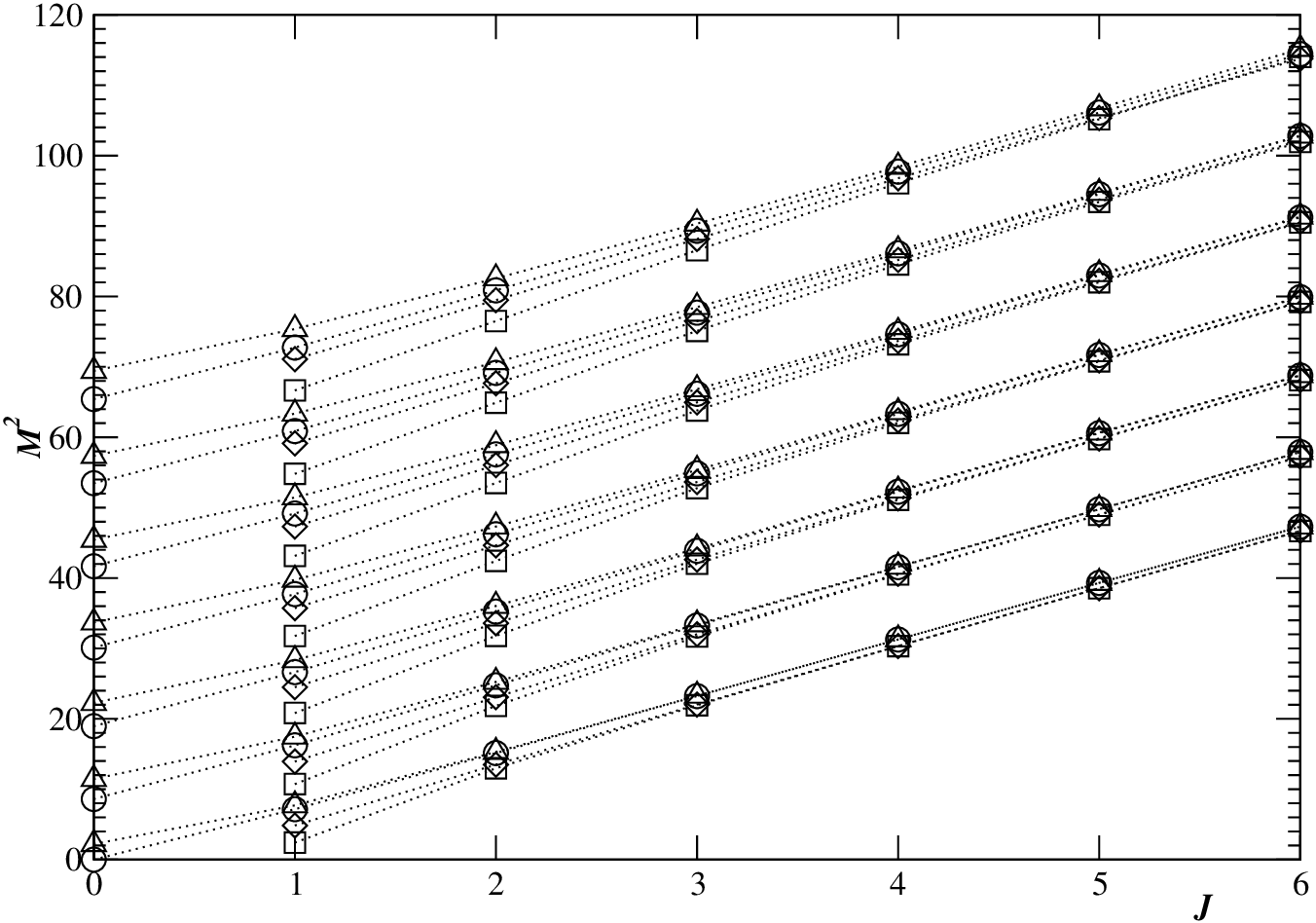}\hfill
\includegraphics[height=5.5cm,clip=]{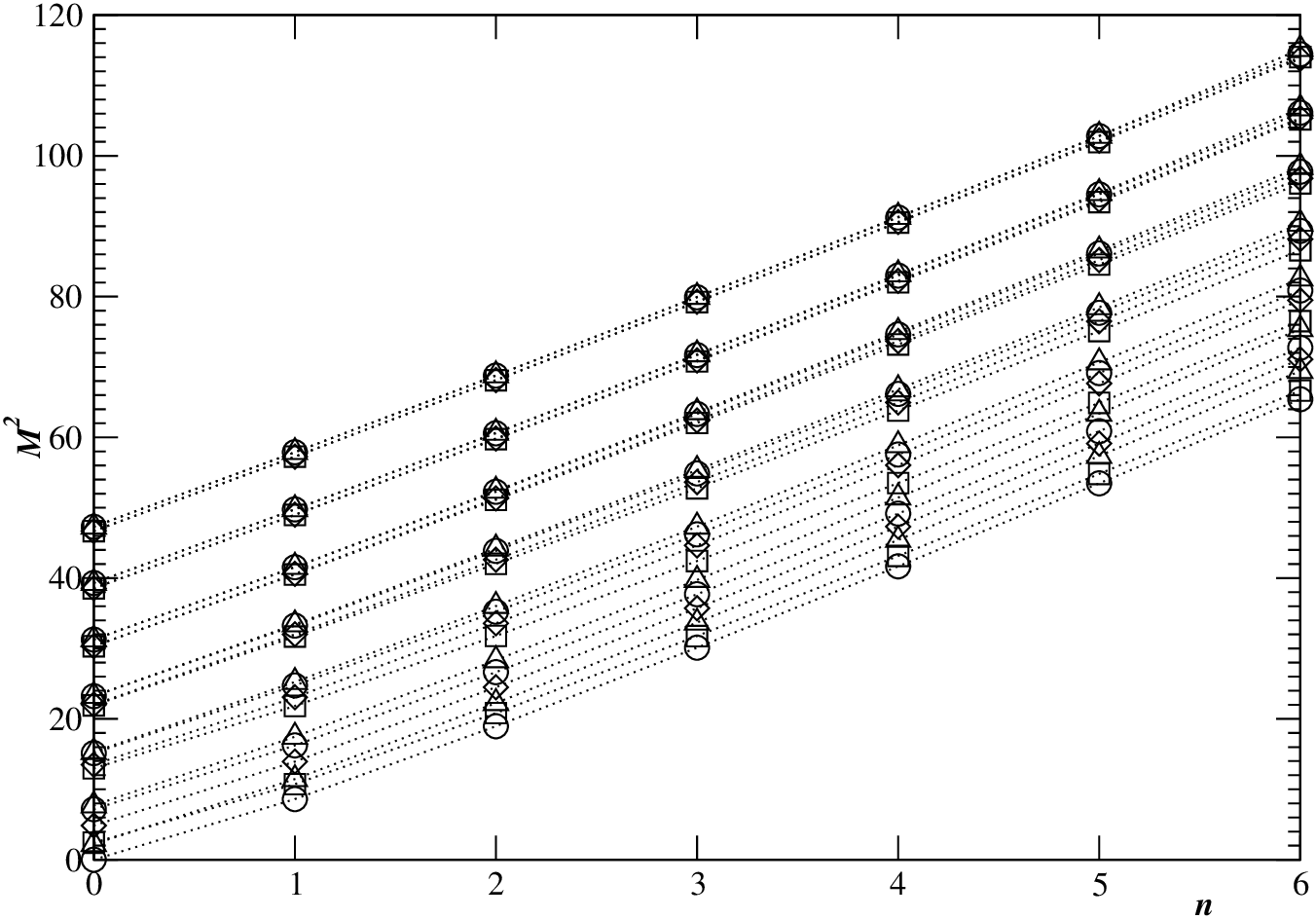}
\caption{Angular (left) and radial (right) Regge trajectories for isovector mesons.
Mesons of the chiral multiplet $(1/2,1/2)_a$ are indicated by circles,
of $(1/2,1/2)_b$ by triangles, and of $(0,1)\oplus(1,0)$
by squares ($J^{++}$ and $J^{--}$ for even and odd $J$, respectively)
and diamonds ($J^{--}$ and $J^{++}$ for even and odd $J$, respectively).
}
\end{figure}

\paragraph*{Acknowledgment:}
The results presented in this contribution were obtained in collaboration with
R. Alkofer, L.~Ya. Glozman, M. Kloker, and A. Krassnigg.  
I am grateful to the organizers for providing financial support
for my participation at the workshop.

\end{document}